\documentclass[prl,floatfix,showpacs,amsmat,twocolumn]{revtex4}
\usepackage{amssymb}
\usepackage{amsmath,bbm}
\usepackage{graphicx}

\def\ketbra #1 #2 {\langle #1\vert #2 \rangle}

\newtheorem{lemma}{Lemma}
\newtheorem{theorem}{Theorem}
\newcommand{\be}{\begin{eqnarray}}
\newcommand{\ee}{\end{eqnarray}}
\newcommand{\bea}{\begin{eqnarray}}
\newcommand{\eea}{\end{eqnarray}}
\newcommand{\bma}{\begin{subequations}}
\newcommand{\ema}{\end{subequations}}
\newcommand{\1}{\mathbbm{1}}
\DeclareMathOperator{\trace}{tr} 

\newcommand{\C}{\mathbb{C}}

\renewcommand{\>}{\rangle}
\newcommand{\<}{\langle}
\newcommand{\vs}{\vspace*{4pt}}
\begin{document}

\title{String order and symmetries in quantum spin lattices}

\author{D. P\'{e}rez-Garc\'{\i}a$^1$, M.M. Wolf$^2$, M. Sanz$^2$, F. Verstraete$^3$, J.I. Cirac$^2$}
\affiliation{$1$ Departamento de Analisis Matematico. Universidad
Complutense de Madrid, 28040 Madrid, Spain.\\
$2$ Max-Planck Institut f\"ur Quantenoptik,
Hans-Kopfermann-Str. 1, Garching, D-85748, Germany \\
$3$Fakult\"{a}t f\"{u}r Physik, Universit\"{a}t Wien, Boltzmanngasse
5, A-1090 Wien, Austria.}

\date{\today}

\begin{abstract}
We show that the existence of string order in a given quantum
state is intimately related to the presence of a local symmetry by
proving that both concepts are equivalent within the framework of
finitely correlated states. Once this connection is established,
we provide a complete characterization of local symmetries in
these states. The results allow to understand in a straightforward
way many of the properties of string order parameters, like their
robustness/fragility under perturbations and their typical
disappearance beyond strictly one-dimensional lattices. We propose
and discuss an alternative definition, ideally suited for
detecting phase transitions, and generalizations to two and more
spatial dimensions.
\end{abstract}

\maketitle


Order parameters play a crucial role in describing the different
phases of matter. However, there exist some phases which, despite
displaying very intriguing features, are not amenable of such a
description. In some cases it is nevertheless possible to
introduce more sophisticated quantities that are able to
characterize those phases. A paradigmatic example is given by the
string order parameter (SOP), which reveals the appearance of a
hidden order (so--called string order, SO) in certain spin systems
\cite{SOP-history-1,SOP-history-2,ladder}. This quantity can be
expressed as an expectation value of some non--local operator, and
the appearance of SO is highlighted by a non--vanishing value in
the thermodynamical limit. Despite its importance, we still do not
have a systematic characterization of its properties. It is not
clear under which conditions SO appears in a 1D  system, for which
kind of small perturbations respecting the gap it is robust
\cite{fragility}, or when it can be used to detect a quantum phase
transition. Apart from that, it seems that the SO looses some of
its desirable properties beyond strictly one-dimensional systems
\cite{ladder,fragility}.

In this work we clarify all those questions for \emph{finitely
correlated states} (FCS) \cite{FaNaWe}, i.e., \emph{matrix product
states} (MPS) \cite{MPS-mother} on infinite chains. The relevance
of these states relies on the fact that every quantum state of a
finite system has an exact MPS representation
\cite{MPS-mother,Vidal} and that ground states of 1D short-range
interactions can be efficiently approximated within this class
\cite{approx}. In this framework we will show that the appearance
of SO is intimately related to the existence of symmetries, which
explains how it can be used to detect quantum phase transitions.
We propose another parameter which better recognizes this
appearance, since it does not have some of the shortcomings of the
SOP. We also provide a natural generalization of SO to  higher
dimensional lattices (membrane order) which retains all the
desired properties. Finally, we give several examples displaying a
large variety of phenomena. \vs

\textbf{1D Chains: String order.} We will mostly consider infinite
 chains of identical spin-$S$ particles in a translationally invariant
state $\Psi$. We say that the state $\Psi$ has string order if
there exist a local unitary $u\ne \1$ and local operators $x,y$
(which can be taken hermitian) such that
 \bea
 \lim_{N\to\infty} |S_N(x,y,u,\Psi)|&>&0,\\
 \label{SOP}
 S_N(x,y,u,\Psi)&:=&\langle\Psi|
 x\otimes u^{\otimes N} \otimes y |\Psi\rangle.
 \eea
Later on we will introduce alternative quantities which will
extend this definition \footnote{In the context of MPS one finds
an alternative but equivalent definition for string order
\cite{Martin-Delgado} involving a finite chain with access to the
virtual levels in the endpoints. }.

Matrix product states of $L$ spins in a translationally invariant
state have the form
 \be
 \label{MPS}
 |\Psi_L\rangle = \sum_{n_1\ldots n_L=-S}^S
 {\rm tr}(A_{n_1}\ldots A_{n_L}) \;|n_1,\ldots,n_n\rangle,
 \ee
where the $A$'s are $D\times D$ matrices. We will write
$\Psi_\infty$ for the thermodynamic limit meaning that
$L\to\infty$ is taken after the expectation value. In this limit
the states are known as finitely correlated states
(FCS)\cite{FaNaWe}---the subject of our studies. Most of the
properties of these states are encoded in a linear map defined as
 \be
 \label{MapE}
 {\cal E}(X)=\sum_{n=-S}^{S} A_n X A_n^\dagger.
 \ee
The matrices $A$ can always be chosen such that
 \be
 \label{fixed}
 {\cal E}(\1)=\1, \quad {\cal E}^*(\Lambda)=\Lambda,
 \ee
where ${\cal E}^*$ denotes the map which is obtained by
interchanging $A_n\leftrightarrow A_n^\dagger$ in (\ref{MapE}),
$\Lambda\ge 0$ and tr$(\Lambda)=1$. Thus, ${\cal E}$ (${\cal
E}^*$) is a unital (trace--preserving) completely positive map,
i.e., a \emph{quantum channel}, and has an eigenvalue equal to 1.
A FCS is a pure state iff $\Lambda>0$ and $\cal E$ has only one
eigenvalue of modulus one. Since there is a unique decomposition
\cite{FaNaWe} of mixed FCS into pure ones and a mixed FCS has SO
iff one of its pure components has it, we will in the following
restrict to pure FCS. Note that all of them are unique ground
states of gapped finite-range interactions. \vs

\textbf{String order and finitely correlated states.} For any
unitary $u$ the SOP (\ref{SOP}) of a FCS is most easily expressed
by introducing a map
 \be
 \label{EU}
 {\cal E}_u(X) := \sum_{n,n'} \langle n'|u|n\rangle  A_{n} X
 A_{n'}^\dagger=\sum_{j} e^{i\theta_j} \tilde A_j X \tilde
 A_j^\dagger,
 \ee
where $\tilde A_j = \sum_n \langle \tilde j|n\rangle A_n$, and
$u=\sum_j e^{i\theta_j} |\tilde j\rangle\langle \tilde j|$. Then
\cite{MPS-mother}
 \be
 \label{SOPMP}
 S_N(x,y,u,\Psi_\infty) = {\rm tr}[\Lambda {\cal E}_x {\cal
 E}_u^N
 {\cal E}_y (\1)],
 \ee
where ${\cal E}_{x,y}$ are defined analogous to ${\cal E}_u$ in
Eq.(\ref{EU}).

The following Lemma studies the spectral radius $\rho$ of ${\cal
E}_u$ which is crucial for Eq.(\ref{SOPMP}) due to the limit
$N\rightarrow\infty$.

\begin{lemma} $\rho({\cal E}_u)\le 1$ with equality iff there exists a unitary $V$
and $\theta\in[0,2\pi)$ such that
 \be
 \label{CondSOP2}
V^\dagger \tilde A_j = e^{i(\theta-\theta_j)}  \tilde A_j
V^\dagger.
 \ee
${\cal E}_u$ has at most one eigenvalue of modulus 1.
\end{lemma}
{\em Proof:} Let us consider an eigenvector, $V$, of ${\cal E}_u$
with eigenvalue $\lambda$, i.e. ${\cal E}_u(V)=\lambda V$.
Multiplying from the right by $\Lambda V^\dagger$ and taking
traces, we obtain
 \bea
 && |\lambda|{\rm tr}(V\Lambda V^\dagger)=\left|\sum_j e^{i\theta_j}{\rm
 tr}(\tilde A_jV\tilde A_j^{\dagger}\Lambda V^\dagger)\right| \nonumber\\
 &\le& \left[\sum_j {\rm
 tr}(V\tilde A_j^{\dagger}\Lambda \tilde A_j V^\dagger)\right]^{1/2}
 \left[\sum_j {\rm tr}(\tilde A_j^{\dagger}V\Lambda V^\dagger\tilde A_j
 )\right]^{1/2} \nonumber\\
 &=&{\rm tr}(V\Lambda V^\dagger),
 \label{ineq}
 \eea
where we have used Cauchy-Schwarz inequality and (\ref{fixed}).
Since $\Lambda>0$, ${\rm tr}(V\Lambda V^\dagger)>0$ and thus
$|\lambda|\le 1$ as stated. Now, if condition (\ref{CondSOP2}) is
fulfilled one can readily see that $e^{i\theta}$ is an eigenvalue
of ${\cal E}_u$ by using Eq.\ (\ref{EU}), and thus $\rho({\cal
E}_u)= 1$. On the contrary, if $|\lambda|=1$, then the inequality
in (\ref{ineq}) has to become an equality, and thus $\alpha
e^{i\theta_j}\Lambda^{1/2} V^\dagger \tilde A_j =\Lambda^{1/2}
\tilde A_j V^\dagger$. Multiplying by the adjoint expression,
taking traces, summing in $j$, and using again (\ref{fixed}) one
obtains that $|\alpha|=1$, i.e. $\alpha=e^{-i\theta}$. Since
$\Lambda$ is invertible we obtain (\ref{CondSOP2}). This also
implies that
 \be
 {\cal E}(V^\dagger V)=V^\dagger\sum_j \tilde A^j \tilde A^{j\dagger}
 V=V^{\dagger} V,
 \ee
where we have used (\ref{fixed}). Since $\1$ is
the only fixed point of ${\cal E}$, we get $V^\dagger V=\1$.
Moreover, suppose that ${\cal E}_u$ has two eigenvectors, $V,V'$
with eigenvalues $e^{i\theta},e^{i\theta'}$, respectively. Then,
using (\ref{ineq}) we have
 \be
 {\cal E}(V^\dagger V')=\sum_j \tilde A_jV^\dagger V' \tilde
 A_j^\dagger=e^{i(\theta'-\theta)} V^\dagger V',
 \ee
such that the same argument gives $V=V'$ and $\theta=\theta'$
$\square$.

Now we can specify the conditions required for SO. First,
$\rho({\cal E}_u)=1$ since otherwise $S_N$ will decay
exponentially with $N$. Using Lemma 1 we know that the eigenvalue
$\lambda$ of magnitude 1 is not degenerate, so let us denote by
$V$ and $Y$ the corresponding right and left eigenvectors, i.e.
${\cal E}_u(V)=\lambda V$, ${\cal E}^*_u(Y)=\lambda Y$, where
${\cal E}^*_u$ is again given by expression (\ref{EU}) but
interchanging $A_n\leftrightarrow A_n^\dagger$. We have
$\lim_{N\to\infty}S_N(x,y,u,\Psi_\infty) = {\rm tr}[Y {\cal
E}_y(\1)] {\rm tr}[\Lambda {\cal E}_x(V)]$. Writing ${\cal
E}_u^*(Y)V=\lambda YV$ and using (\ref{CondSOP2}) we arrive to the
conclusion that $Y=\Lambda V^\dagger$. Thus, the conditions for
the SOP not to vanish are: (i) $\rho({\cal E}_u)=1$; (ii) ${\rm
tr}[\Lambda V^\dagger {\cal E}_y(\1)]$, ${\rm tr}[\Lambda {\cal
E}_x(V)]\ne 0$.

We may ask ourselves if the condition $\rho({\cal E}_u)=1$ is
sufficient to have SO, i.e. if there are always two operators $x$
and $y$ such that the other conditions are fulfilled. To answer
this question we notice that ${\rm tr}[\Lambda V^\dagger {\cal
E}_y(\1)]^\ast={\rm tr}[\Lambda {\cal E}_z(V)]$, where $z=\tilde u
y$ and $\tilde u=\sum_j e^{i(\theta_j-\theta)}|\tilde
j\rangle\langle \tilde j|$. Thus, we can always choose
$y=x^\dagger \tilde u$ so that condition (ii) above is simplified
to ${\rm tr}[\Lambda {\cal E}_x(V)]\ne 0$. It is clear that it
suffices that ${\rm tr}(V\Lambda A_n A_m^\dagger)\ne 0$ for some
$n,m$ since then we can simply choose $x=|n\rangle\langle m|$. We
conclude that

\begin{theorem} For a pure FCS there exists SO iff there exist a
unitary $\tilde u\ne \1$, $V$, and $n,m$ such that
 \be
 \label{Th1}
 {\cal E}_{\tilde u}(V)=V,\quad {\rm tr}(V\Lambda A_n A_m^\dagger)\ne 0.
 \ee
\end{theorem}

Now, we show that the second condition can be dropped in two
situations. First, if $x$ and $y$ in (\ref{SOP}) are products of
observables acting on $D^2$ spins. The reason is that the set
$S_D:=\text{span}\{A_{n_1}\ldots A_{n_D} A_{m_D}^\dagger\ldots
A_{m_1}^\dagger\}$ spans the set of $D\times D$ matrices, so that
there is always a linear combination $X$ of these matrices for
which ${\rm tr}(V\Lambda X)\ne 0$. To see that this set is
complete note first that $S_{m-1}\subseteq S_m$ since
$\sum_{n_D}A_{n_1}\ldots A_{n_D} A_{n_D}^\dagger\ldots
A_{m_1}^\dagger=A_{n_1}\ldots A_{n_{D-1}}
A_{m_{D.1}}^\dagger\ldots A_{m_1}^\dagger$. This inclusion must be
strict unless $m> D^2$ since $S_{m-1}=S_m$ implies $S_m=S_{m+1}$
and for a sufficient large $N$ the set $\{A_{n_1}\ldots A_{n_N}\}$
must span the entire space of matrices \cite{MPS-mother}. Another
situation is the one in which there exists a continuous group of
unitaries $V$ fulfilling the first condition, i.e., we can
parametrize $V=e^{i\phi H}$. Then, we can always choose $\phi$
sufficiently small such that ${\rm tr}(V\Lambda)\ne 0$ and this
suffices since we have (\ref{fixed}). \vs

\textbf{Symmetries in finitely correlated states.} We say that a
state has a local symmetry if there is a unitary $u\neq\1$ such
that
 $$u\otimes \cdots \otimes
u |\Psi\>=|\Psi\>.$$ This formally means that for every $N$-site
reduced density operator $\varrho$ we have $ u^{\otimes N}\varrho
u^{\dagger \; \otimes N}=\varrho$.

For FCS the condition $\rho({\cal E}_u)=1$ is not only equivalent
to having SO---as we saw in the previous section---it is also
equivalent to the presence of a local symmetry:

\begin{theorem} A pure FCS has a local symmetry $u$ iff $\rho({\cal
E}_u)=1$.\end{theorem}

{\it Proof:} If $\rho({\cal E}_u)=1$ the result is a direct
consequence of Lemma 1 which implies that $V^\dagger \Lambda
V=\Lambda$. The converse follows from the fact that
$$\frac{1}{D^2}\le \trace(\varrho^2)=\trace[\varrho u^{\otimes N}\varrho
u^{\dagger \otimes N}]={\rm tr}\Big[L({\cal E}_u\otimes {\cal
E}_{u^\dagger})^N(R)\Big],$$ for some $L,R$ which are independent
of $N$. $\square$

This theorem, together with Lemma 1 provides a complete
characterization of FCS with local symmetries. We note that an
analogous statement can be found (though without proof) in
Ref.\cite{FaNaWe}.

Equivalent criteria for the existence of a local symmetry can be
given in terms of the isometry $B:=\sum_j |j\rangle A_j$ as well
as for the $D^2\times D^2$ matrix $E:=\sum_j A_j\otimes \bar A_j$.
In all cases $V$ and $u$ are elements of two unitary
representations of a symmetry group (different from the identity):
\begin{itemize}
    \item Condition C1: $(u\otimes \1)B=(\1\otimes V)BV^\dagger$.
    \item Condition C2: ${\cal E}$ is
covariant, i.e., for all $X$, ${\cal E}(V X V^\dagger)=V{\cal
E}(X)V^\dagger$.
    \item Condition C3: $[E, (V\otimes \bar V)]=0$.
\end{itemize}
 C1 can
be obtained from Lemma 1 by using the spectral decomposition of
$u$. C2 and C3 are the same if we use $\langle
k,l|E|i,j\rangle=\langle k|{\cal E}(|i\rangle\langle
j|)|l\rangle$. It is also clear that Lemma 1 implies C2. Finally,
if C3 is fulfilled, then we have that $VA_j V^\dagger$ are also
Kraus operators of the map ${\cal E}$. Since all Kraus
decompositions are related by a unitary matrix, say $u$, we have
that $VA_j V^\dagger = \sum_n \langle n|u|j\rangle A_n$ implying
the condition of Lemma 1.

Now, we can use C3 to derive a criterion for the existence of a
{\em continuous symmetry} where $V=e^{i\phi H}$. Expanding in
first order in $\phi$ we obtain
 \be
 M(H):= [E,H\otimes\1  -\1\otimes \bar H]=0.
 \ee
Since $M$ is a linear map on the space of hermitian matrices, we
have that: a pure FCS has a local continuous symmetry iff $M$ has
a non-trivial kernel.

Similar criteria can be given for discrete symmetries, e.g.,
$\mathbb{Z}_2$ symmetry where the $A$'s are either block diagonal
or block off-diagonal \cite{forthcoming}. \vs

\textbf{Alternative definitions.} One can understand the
importance of the SOP to detect quantum phase transition in terms
of its relation to local symmetries. If we have a Hamiltonian with
certain local symmetry and its ground state is unique, then there
may be SO. If we change the parameters of the Hamiltonian but
keeping the symmetry until the gap closes, then the ground state
will be degenerate and the symmetry may be broken \footnote{If one
perturbs the Hamiltonian without keeping the symmetry, it is clear
from our picture that the string order will vanish, even if the
gap does not close. This explains the fragility of the SO (cf.
\cite{fragility})}. Thus, the SO may disappear at that point,
indicating the presence of the transition. Note that due to the
possible choices of the operators $x$ and $y$ in the definition
(\ref{SOP}), it may happen that the SOP for a particular choice
vanishes even if there still is a symmetry. In order to avoid
this, one may look at the quantity
 \be
 \label{RL}
 R_L(u):=\langle\Psi_L|u^{\otimes L}|\Psi_L\rangle,
 \ee
 where $|\Psi_L\>$ is the ground state of the Hamiltonian acting on $L$ sites with periodic boundary conditions (Eq. (\ref{MPS}) in the case of MPS), which is
indeed directly related to the existence of a symmetry. In fact,
$R_\infty(u)=\lim_L R_L(u)$ can only vanish if the gap is closed,
and thus it is ideally suited to study the presence of
transitions. Note also that it can be straightforwardly determined
from numerical algorithms based on MPS \cite{MPS-numerics}. \vs

{\bf Example 1:} {\em AKLT state}. It is instructive to revise the
appearance of SO in the ground state of the AKLT model
\cite{AKLT}. For that state we have $S=1$, $A^0=\sigma_z/\sqrt{3}$
and $A^{1,-1}=\sqrt{2/3}\sigma_\pm$ where the sigmas denote Pauli
matrices. One finds $\Lambda=\1$, and taking $u=e^{i\pi S_z}$ we
obtain $V=\sigma_z$ and we can take $x=y=S_z$, so that
$S_L(x,y,u,\Psi_\infty)=-4/9$ \cite{SOP-history-1}. Note that the
AKLT state has  $SU(2)$ symmetry so that obviously $R_L(u)=1$. \vs

{\bf Example 2:} {\em Cluster state}. We have \cite{MPS-mother}
 \be
 A^0 = \frac{1}{\sqrt{2}} \left(\begin{array}{cc} 1 & 1 \\ 0 & 0 \end{array}\right)\quad
 A^1 = \frac{1}{\sqrt{2}}\left(\begin{array}{cc} 0 & 0 \\ 1 & -1 \end{array}\right).
 \ee
This has the symmetry induced by $u=-\sigma_x$. We have
$V=\sigma_y$, and one can readily see that ${\rm tr}(V\Lambda A_n
A_j^\dagger)= 0$. Thus, there is no SO. However, if we take two
particles, we can choose $x=\sigma_z\otimes\sigma_y$ and
$y=\sigma_y\otimes\sigma_z$, so that the SOP is one. In general,
we have $R_L(u)=1$ as expected. \vs

\textbf{2D Systems: Membrane order.}
Since the existence of a local symmetry, and the possibility of determining it numerically via the
 quantity $R_L$ defined in (\ref{RL}), seems to be an appropriate definition of SO in 1D systems,
  we will now try to extend it to 2D. In this case there is more freedom in the choice
  of locations for the local unitaries $u$. First we can let them act on the whole lattice. If this leaves the state invariant (or the respective $R_L$ is not vanishing) we will say that
we have membrane order (MO). We can also put them as a string of
operators, in which case we will talk about SO \footnote{In this
case one can obtain again the equivalence with the 'classical'
definition of string order. In order to see this, it is enough to
invoke the injectivity condition defined in \cite{PEPS-2}.}, or
even in more sophisticated configurations, as a band of operators
(BO). Note that the MO so defined shares all the desired
properties for the SO in 1D and thus provides a natural
generalization to higher dimensions. In particular, it should not
exhibit spontaneous breakdown when switching on couplings in the
second dimension as pointed out in \cite{fragility} for the SO.

\begin{figure}
  \includegraphics[width=8cm]{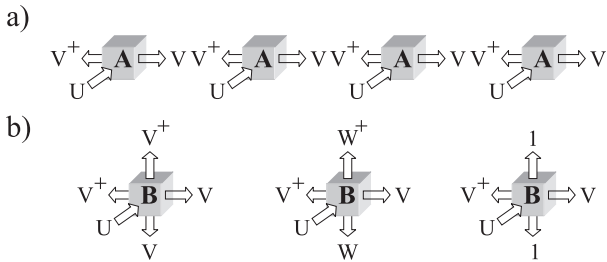}\\
 \caption{(a) Symmetries in FCS/MPS. The tensor $A$ has
 three indices, one corresponding to the physical spin (pointing
 in) and two for the virtual systems (pointing left and
 right). Applying $U$ to the physical index is equivalent to
 applying $V$ to the virtual ones. Since the tensor are contracted
 in a line the unitaries $V$
 and $V^\dagger$ cancel and thus the state
 does not change. (b) 2D generalization (square lattice). The tensor $B$ has one
 physical and four virtual indices. When applying $U$ to the
 former, we can have different effects on the virtual system which
 will, after contraction, leave the state invariant. The example on the right allows for SO.}
 \label{Fig1}
 \end{figure}

To gain more inside we will consider the generalization of MPS/FCS
called \emph{projected entangled pair states} (PEPS)
\cite{PEPS-1,PEPS-2} where the matrices $A_k$ are replaced by
tensors $B_k$ whose degree depends on the geometry of the lattice.
For these states local symmetries can arise in a similar way as in
the 1D case  (Fig.\ref{Fig1}). We interpret each tensor as an
operator $B=\sum_s|s\rangle\langle\phi_s|:(H_D)^{\otimes 4}\to
H_S$, where $H_S$ ($H_D$) is the Hilbert space corresponding to
the physical (virtual) spins. The PEPS exhibits a local symmetry
if there exist unitaries $V_k$ and $U\neq\1$ with
\begin{equation}\label{sym-PEPS}
U B = B V_1\otimes ... \otimes V_4,\end{equation} such that when
contracting the indices of $B$ to create the state $\Psi$
(analogous to the matrices $A$ in Eq.(\ref{MPS})) the $V$'s cancel
[see Fig.\ \ref{Fig1}]. Thus, we will have $\langle
\Psi|U^{\otimes N}|\Psi\rangle=|\Psi\rangle$ and therefore MO.
Note that we have the possibility of having different $V$'s, in
contrast to what happens in 1D. This structure also allows us to
understand the (dis-) appearance of SO or BO. However, in the 2D
case, the connection between the existence of a local symmetry and
Eq. (\ref{sym-PEPS}) is less straight and will be analyzed in
detail elsewhere \cite{forthcoming}.\vs

{\bf Example 3:} {\em AKLT state} \cite{AKLT}. In this case $S=2$,
$|\phi_s\rangle=(\sigma_y\otimes\sigma_y\otimes 1\otimes 1)|\bar
\phi_s\rangle$ and $\phi_s$ is an orthonormal basis of the
symmetric subspace of $(\C^2)^{\otimes 4}$. $U$ and $V$ correspond
to a 5 and 2--dimensional representation of $SU(2)$, and thus
fulfill the condition of MO. One can prove that there is no SO
 even for a simple ladder formed by two chains.
\vs

{\bf Example 4:} {\em Cluster state} \cite{cluster}. Here, $S=1/2$
and $|\phi_{-1/2}\rangle=|++00\rangle$,
$|\phi_{1/2}\rangle=|--11\rangle$ with
$|\pm\rangle=|0\rangle\pm|1\rangle$. One can take $U=\sigma_x$,
$V_u=V_r=\sigma_x$ and $V_d=V_l=\sigma_z$. But one can also take
$U=\sigma_z$, three of the $V$'s equal to $\1$ and decide either
$V_u=\sigma_z$, $V_r=\sigma_z$, $V_d=\sigma_x$ or $V_l=\sigma_x$.
As a consequence, there is MO. One can readily show that there is
no SO. However still there is a BO, in the sense that this state
is an eigenstate of the operator obtained by applying unitaries to
three consecutive lines ($\sigma_z$ to the first and the third and
$\sigma_x$ to the middel one). \vs

{\bf Example 5:} {\em Toric code} \cite{Kitaev}. This is a state
with $S=1/2$ and we have alternative tensors in the A and B
sublattices, In this case we have alternating projectors
$B=|0\>\<\Phi_+\Phi_+|_{uldr}+|1\>\<\Phi_{-}\Phi_{-}|_{uldr}$ and
$B'=|0\>\<\Phi_+\Phi_+|_{urld}+|1\>\<\Phi_{-}\Phi_{-}|_{urld}$,
resp. One can take $U=\sigma_x$, $V_{l}=V_{r}=\sigma_z$ and
$V_{u}=V_{d}=\1$. Since there is no unitary to be cancelled in the
up and down positions, it has SO (and also MO).\vs


\textbf{Conclusion.} We have shown that the existence of string
order is, in the framework of FCS, equivalent to the existence of
a local symmetry. This gives a direct explanation of many of its
intriguing features, like its robustness/fragility under
perturbations or the capability of detecting phase transitions.
Though the FCS case provides evidence for the generality of this
equivalence it remains an open problem to extend this beyond FCS,
e.g., to all ground states of gapped local Hamiltonains. We have
characterized the existence of local symmetries in FCS, related it
to intertwining isometries and covariant channels (C1 and C2), and
sketched the case of 2D systems through PEPS. The list of examples
can easily be extended and shows, for instance, in a simple way
the presence of string order in topological ordered states, as it
is illustrated in Kitaev's toric code.

The present work is an example that MPS/FCS and PEPS are not only
useful for numerical algorithms, but also to prove and clarify
interesting statements.

The obtained results shed light on the role of symmetries in spin
systems also in other contexts like Lieb-Schultz-Mattis type
theorems \cite{Lieb}. The respective relation between
integer/half-integer spins and irreducible/reducible
representations will be discussed in a forthcoming paper
\cite{forthcoming} together with a more detailed investigation of
the conditions for symmetry.\vs

The authors thank A.Rosch and E.Altman for discussions. Portions
of this work were done at the Workshop on Tensor network methods
and entanglement at the Erwin Schr\"{o}dinger Institute. This work
has been supported by the EU project SCALA, the DFG (FOR 635, MAP
and NIM) and the Spanish grant MTM2005-00082.


\begin{thebibliography}{99}
\bibitem{SOP-history-1} M. den Nijs and K. Rommelse, Phys. Rev. B {\bf 40}, 4709
(1989).
\bibitem{SOP-history-2} E. G. Dalla Torre, E. Berg, E. Altman, Phys. Rev. Lett. {\bf 97}, 260401 (2006); F. Anfuso, A. Rosch Phys. Rev. B {\bf 75}, 144420 (2007).
\bibitem{ladder}   Eugene H. Kim, G. Fath, J. Solyom, D. J. Scalapino Phys. Rev. B {\bf 62}, 14965 (2000); S. Todo, M. Matsumoto, C. Yasuda, H. Takayama, Phys. Rev. B {\bf 64}, 224412 (2001).
\bibitem{fragility} F. Anfuso, A. Rosch Phys. Rev. B {\bf 76}, 085124 (2007).
\bibitem{MPS-mother}  D.
P\'erez-Garc\'{\i}a, F. Verstraete, M.M. Wolf,  J.I. Cirac, Quant.
Inf. Comp. {\bf 7}, 401 (2007).
\bibitem{FaNaWe}  M. Fannes,
B. Nachtergaele and R. F. Werner, Commun. Math. Phys.
\textbf{144}, 443-490 (1992).
\bibitem{Vidal} G. Vidal, Phys. Rev. Lett. {\bf 91}, 147902
(2003).
\bibitem{Martin-Delgado}  F. Verstraete, M.A. Martin-Delgado, J.I. Cirac, Phys. Rev. Lett. {\bf 92}, 087201 (2004).
\bibitem{approx} F. Verstraete, J.I. Cirac, Phys. Rev. B {\bf 73}, 094423 (2006); M. B. Hastings J. Stat. Mech. (2007) P08024.
\bibitem{MPS-numerics} F. Verstraete, D. Porras, J.I. Cirac, Phys. Rev. Lett  {\bf 93}, 227205 (2004).
\bibitem{AKLT} I. Affleck, T. Kennedy, E. H. Lieb and H.
Tasaki, Commun. Math. Phys. \textbf{115}, 477 (1988).
\bibitem{PEPS-1} F. Verstraete, J.I. Cirac, arXiv:cond-mat/0407066.
\bibitem{PEPS-2}  D.
P\'erez-Garc\'{\i}a, F. Verstraete, M.M. Wolf,  J.I. Cirac, arXiv:0707.2260.
\bibitem{forthcoming} In preparation.
\bibitem{cluster} R. Raussendorf, H.J. Briegel, Phys. Rev. Lett. 86,
5188 (2001); F. Verstraete, J.I. Cirac, Phys. Rev. A
\textbf{70}, 060302(R) (2004).
\bibitem{Kitaev} A. Yu. Kitaev, Annals Phys. \textbf{303}, 2 (2003).
\bibitem{Lieb}  E. Lieb, T. Schultz, and D. Mattis, Ann. Phys.
{\bf 16} (1961), 407–466; M. B. Hastings, Phys.Rev. B {\bf 69} (2004) 104431;  B. Nachtergaele, R. Sims, Commun. Math. Phys. {\bf 276} (2007) 437--472.

\end{thebibliography}
\end{document}